# Estimation of the Effect of Carbon Tax Implementation on Household Income Distribution in Indonesia: Quantitative Analysis with Miyazawa Input- Output Approach

**Syahrituah Siregar**
**University of Lambung Mangkurat**
https://feb.ulm.ac.id/id/staf-akademik/
**Correspondence: syahrituahsiregar.iesp@ulm.ac.id**

## ABSTRACT

Climate change is a global challenge caused by greenhouse gas emissions from fossil fuel use. Indonesia, as a developing country, faces major challenges in implementing carbon tax policies to reduce emissions, especially related to their regressive impacts on low-income households. Currently, there is little in-depth research on how carbon tax policies impact household income distribution in Indonesia. This study uses a quantitative approach with the Input- Output model to analyze the impact of carbon tax on household income based on 10 income groups, both in urban and rural areas. The results show that carbon tax policies have a regressive impact, where low-income households bear a proportionally greater burden. Household income in **Class - 10** decreased by IDR 19,144.85 million in urban areas and IDR 8,819.13 million in rural areas, while households in **Class - 1** decreased by IDR 954.23 million. Therefore, mitigation policies such as cross subsidies are needed to reduce the impact on vulnerable groups. These findings are important for policy makers in formulating fair and effective fiscal policies, as well as ensuring social justice in the context of sustainable development. This study has limitations in the scope of analysis of long-term energy consumption behavior and certain sectors, so further research is needed to deepen these aspects.

**Keywords:** *regressive impact; income distribution; carbon tax; sustainable development; household*

## INTRODUCTION

Global climate change has become a major challenge due to the increase in greenhouse gas emissions, especially carbon dioxide ($CO_2$) from the burning of fossil fuels such as oil, gas, and coal. $CO_2$ emissions accelerate global temperature rise, threatening ecosystems and human societies. To limit the temperature, rise to 1.5°C, significant and urgent emission reductions are needed. However, despite many countries committing to reduce emissions, global emissions still increased by 1.1% in 2023, primarily due to energy needs in developing countries (IPCC, 2022, 2023).

The impact of this increase in emissions is very wide-ranging, including rising sea levels, melting polar ice, and extreme weather changes that cause disasters such as floods, droughts, and heatwaves. Emissions from the energy sector, particularly fossil fuels, contribute the most to global greenhouse gases. Countries with the highest emission levels, such as the United States, China, and the European Union, are responsible for more than half of the total global emissions (Rosado & Ritchie, 2022). Therefore, cooperation from these countries is crucial in implementing policies that promote the transition to clean and environmentally friendly energy.

The implementation of carbon taxes in various countries has become an effective policy instrument for reducing $CO_2$ emissions. This policy assigns a price to the carbon produced, so sectors with high fossil fuel intensity must pay a higher cost. Countries that implement carbon pricing are significantly able to reduce their carbon emission growth by up to 2% per year (Best et al., 2020). An increase in the carbon price by one euro per ton of $CO_2$ emissions can reduce emissions by up to 0.3% per year. Countries like Sweden, which implemented a carbon tax since 1991, have reduced carbon emissions by 30% by 2015 compared to if the policy had not been implemented (Martinsson et al., 2024). The implementation of a sustainable carbon tax, although requiring policy adjustments, can have a real impact on emission reductions, especially in countries with adequate economic capacity and infrastructure. The main instrument in carbon tax policy is the imposition of a price on $CO_2$ emissions, either in the form of a direct tax or an emissions trading system. In Canada, a carbon tax is applied to fossil fuels, successfully reducing emissions without harming the economy (The World Bank, 2024a). Countries like Sweden and the UK have also successfully used carbon taxes to reduce emission impacts in the long term (World Economic Forum, 2023). The procedure for implementing carbon tax requires careful design and gradual implementation. The government must consider the sectors being taxed, the rates applied, and the allocation of revenue from this tax. The upstream approach is often chosen because it is more administratively efficient and allows for easier monitoring (C2ES, 2024). Carbon pricing mechanisms should consider the rate of tax escalation to provide a strong signal for industries to invest in clean technology (The World Bank, 2024b).

The evaluation and adjustment process is also important. The government uses economic models to project the impact of carbon taxes on the economy and emissions. Revenue from the carbon tax is often used to fund renewable energy or assist low-income households affected by rising energy prices (The World Bank, 2023). This measure maintains socio-economic stability during the transition to a low-carbon economy. The implementation of carbon tax has significant distributive implications, especially for household income groups. Carbon taxes tend to be regressive because low-income groups are more vulnerable to increases in energy prices (Rosenberg et al., 2018). Furthermore, it has been revealed that several countries have implemented per capita household rebates or income refunds or income tax reductions for low-income households to maintain a balance in the tax burden. Research shows that appropriate income rebate policies can transform the regressive nature of carbon taxes into progressive ones, providing greater benefits to the wider community and achieving the desired environmental goals. Although carbon taxes are considered an effective policy tool for reducing carbon emissions, their effectiveness is often limited due to the ongoing global dependence on fossil fuels. Large investments in fossil fuel infrastructure and economic dependence on fossil energy slow down the energy transition process (World Economic Forum, 2023). Additionally, sudden increases in fuel prices can cause economic disruptions such as temporary unemployment and supply-demand imbalances in the renewable energy sector (The World Bank, 2024b).

Besides carbon dioxide, other greenhouse gas emissions such as methane, which have a greater warming impact in the short term, are often not included in the scope of carbon taxes (The World Bank, 2024a). This explains why many countries also use additional policies such as renewable energy subsidies and environmental regulations to achieve overall emission reduction targets. Dependence on fossil fuels in various economic sectors, both in developed and developing countries, is a major obstacle to the transition to clean energy. High carbon prices can reduce demand for goods that produce emissions, but structural changes in the economy cannot happen quickly (World Economic Forum, 2023). The transition to renewable energy takes time because capital and labor cannot instantly shift to new sectors, causing a short-term imbalance between energy supply and demand (The World Bank, 2024a).

Economic dependence on the fossil fuel sector makes economic restructuring complex. In countries with good infrastructure, disinvestment in fossil fuels is not automatically offset by increased investment in renewable energy, triggering economic uncertainty (Finkelstein Shapiro & Metcalf, 2023). Factors such as the size of fossil fuel reserves, stranded assets, and uncertain climate policy risks slow down the pace of the structural changes needed for a successful energy transition. The implementation of carbon taxes often has a greater impact on low-income households because their proportion of spending on energy is higher compared to middle and upper-income groups. Without effective mitigation policies, they will face rising prices of goods and services, especially energy, which will erode their purchasing power (Columbia University SIPA, 2018). In a scenario with a carbon tax of $50 per ton, analysis shows that the tax burden tends to be regressive unless the proceeds are returned in the form of subsidies or rebates to low-income groups (Boyd et al., 2024; Goulder et al., 2018).

Some countries design compensation policies, such as climate rebates, to return a portion of the revenue to vulnerable households. This can be done through tax credits or direct transfers to poor households, covering a significant portion of the cost increase due to the carbon tax (Stone, 2015). This mechanism is important to maintain socio-economic balance and prevent an increase in income inequality.

The design of revenue recycling policies is an important element in the implementation of carbon taxes. The return of carbon tax revenue to the community in the form of direct transfers or income tax reductions for low-income households aims to balance the impact of rising energy prices due to the carbon tax, especially for vulnerable groups (Shang, 2021). In developed countries, income tax reductions for low-income groups have been proven to reduce regressive impacts and enhance long-term welfare (Rosenberg et al., 2018).

Increased public spending on education and health can also reduce inequality, especially if directed towards low-income households and underdeveloped regions. Green investments in environmentally friendly infrastructure are capable of creating new jobs and improving economic access in sectors that have long been underdeveloped (Shang, 2021). Comprehensively designed income recycling policies can address the potential negative impacts on low-income groups. In Indonesia, an in-depth analysis is needed regarding the impact of carbon tax on household income distribution. Research shows that carbon taxes tend to be regressive, resulting in low-income households bearing a greater proportional tax burden compared to high-income groups (Petra Christi, 2022). Without mitigation policies such as renewable energy subsidies or income compensation programs, income inequality could worsen. Indonesia needs to design appropriate policy schemes to avoid further pressure on the national budget and ensure that fossil fuel subsidies are redirected to renewable energy (Alonso & Kilpatrick, 2022). Efforts to reduce emissions through carbon taxes must be accompanied by significant behavioural changes in the industrial sector. Lessons from other successful countries can help Indonesia design policies that mitigate the negative socio-economic impacts of carbon taxes(Petra Christi, 2022).

This research is important to understand the socio-economic impact of carbon tax policies implemented in various countries, including Indonesia. The regressive impact of carbon tax on low-income households demands a deep understanding of the distribution of this policy's effects. Without comprehensive analysis, carbon tax policies could exacerbate social inequality, contradicting sustainable development goals. The findings of this research are expected to provide comprehensive evidence-based solutions and formulate more equitable and inclusive mitigation policies. This research aims to explore the impact of carbon tax on household income distribution in Indonesia using an Input-Output model. This research also aims to identify the most vulnerable household groups affected by this policy, providing appropriate and fair mitigation policy recommendations. The focus of the research is to analyse the regressive impact of the carbon tax and evaluate the effectiveness of implementable compensation policies. It is expected that this research will make an important contribution to the formulation of fiscal policies that support sustainable development in Indonesia. The hypothesis of this research is that carbon tax has a greater impact on low-income households compared to middle and high-income households. Low-income households allocate a larger proportion of their income to energy expenditures, so the carbon tax policy is expected to significantly increase their expenditure burden. Without adequate mitigation policies such as subsidies or income compensation programs, the regressive impact of the carbon tax will worsen income inequality. Carbon tax policies need to be accompanied by mechanisms to mitigate negative impacts on vulnerable groups in order to achieve balanced social and environmental goals. The conclusion of this analysis will serve as the basis for evaluating the effectiveness of compensation policies and formulating more inclusive policies.

## METHODOLOGY

This research method uses a quantitative approach with a descriptive-analytical design. The analysis was conducted using the Miyazawa Input-Output model to estimate the impact of the carbon tax on household income distribution in Indonesia. The Input-Output table was modified by adding carbon emission levels per sector based on energy consumption estimates. A carbon tax rate of Rp.30,-/kg CO2e is applied as a fiscal intervention shock in the simulation. This model was chosen because it can illustrate the flow of income between sectors and social groups, as well as the impact of fiscal policy on the economic structure. The data used comes from the 2016 Indonesian Input-Output Table, which provides detailed information about transactions between economic sectors. This approach aims to provide an empirical picture of the regressive impact of carbon tax on various income groups. The population of this study includes all households in Indonesia, both urban and rural, divided into 10 income groups. The data sources come from the aggregation of household data in Susenas and secondary data from the Central Statistics Agency (BPS, 2021) as well as the Ministry of Finance (BKF, 2021). Data collection is based on the availability of relevant and up-to-date data.

The main instrument used is the Input-Output model simulation, where changes in the carbon tax variable are incorporated to measure its impact on household income. The research procedure begins with the collection of data from relevant sources, followed by data processing within the Input-Output model. Next, simulations are conducted to calculate the changes in final demand due to the carbon tax, which affects the income of each household group both in urban and rural areas. The simulation results were then analyzed to see the regressive impact of the tax.

## RESULTS AND DISCUSSION

### Results

The research results show that the implementation of carbon tax has a significant impact on the distribution of household income in Indonesia, especially among the lower-middle-income groups. Based on the Input-Output analysis, there is a varying decrease in household income across different income classes, with the low-income group experiencing the greatest impact in terms of percentage reduction in income. The largest nominal income decrease was recorded in the upper-middle-income class, but in relative percentage, it was smaller compared to the lower-income class. The data also reveals that although the absolute impact on higher income groups appears larger, low-income households are more vulnerable to this carbon tax policy. Overall, this study highlights the importance of designing mitigation policies to minimize the regressive impact of carbon taxes on low- and middle-income households.

Table 1. Income Levels and Estimated Rates of Decline Income Based on Class Community Revenue from Carbon Tax in Indonesia (database: IO Table Indonesia 2016)

| No | Income Group | Y1 ( Rp. Million ) | DY ( Rp. Million ) | Y2 ( Rp. Million ) | %DY | %CY |
|---|---|---|---|---|---|---|

| No | Class | | | | | |
|---|---|---|---|---|---|---|
| 1 | Class - 1 | 162,856,815 | 1,639.65 | 162,855,175 | 0.0010068% | 2.18% |
| 2 | Class - 2 | 260,381,265 | 2,592.88 | 260,378,673 | 0.0009958% | 3.45% |
| 3 | Class - 3 | 326,679,028 | 3,239.68 | 326,675,788 | 0.0009917% | 4.31% |
| 4 | Class - 4 | 390.441.241 | 3,857.56 | 390,437,384 | 0.0009880% | 5.14% |
| 5 | Class - 5 | 464,644,203 | 4,583.77 | 464,639,619 | 0.0009865% | 6.10% |
| 6 | Class - 6 | 550,599,409 | 5,420.81 | 550,593,988 | 0.0009845% | 7.22% |
| 7 | Class - 7 | 661,970,128 | 6,507.71 | 661,963,620 | 0.0009831% | 8.66% |
| 8 | Class - 8 | 831.187.331 | 8,146.07 | 831.179.185 | 0.0009801% | 10.85% |
| 9 | Class - 9 | 1,136,537,321 | 11,161.19 | 1,136,526,160 | 0.0009820% | 14.86% |
| 10 | Class - 10 | 2,879,595,098 | 27,963.99 | 2,879,567,134 | 0.0009711% | 37.23% |
| | Total | 7,664,891,840 | 75,113.30 | 7,664,816,726 | 0.0009800% | 100.00% |

*Nominal Impact and Relative Impact*

This research found that the implementation of carbon tax significantly impacts the reduction of household income across all income classes, both in urban and rural areas. In the highest income group of households, namely Class - 10, there was a decrease in income of Rp 19,144.85 million in urban areas, while in rural areas, the decrease in income was recorded at Rp 8,819.13 million. In total, the income decrease in Class - 10 reached Rp 27,963.99 million. Although in percentage terms, this decline seems small, at 0.00142300%, its nominal impact is quite significant, especially on high-income households in urban areas.

In the Class - 1 group, which consists of the lowest-income households in urban areas, the initial income was recorded at Rp 94,750.902 million. The implementation of the carbon tax caused a decrease estimated at Rp 954,23409 million, resulting in a final income of Rp 94,749,947 million. The percentage decrease in this group is 0.00100710%, which indicates that although the nominal decrease is smaller compared to the high-income group, its proportional impact is greater on the low-income group.

The middle group, such as Class - 5, also experienced a significant decline in income. The initial income in this urban group was recorded at Rp 280,984.146 million, and if the carbon tax is applied, the income decrease reaches Rp 2,767.98820 million. The final income for this group became Rp 280,981.378 million, with a percentage decrease of 0.00098500%. Although this percentage is smaller compared to the low-income group, the nominal decrease remains significant.

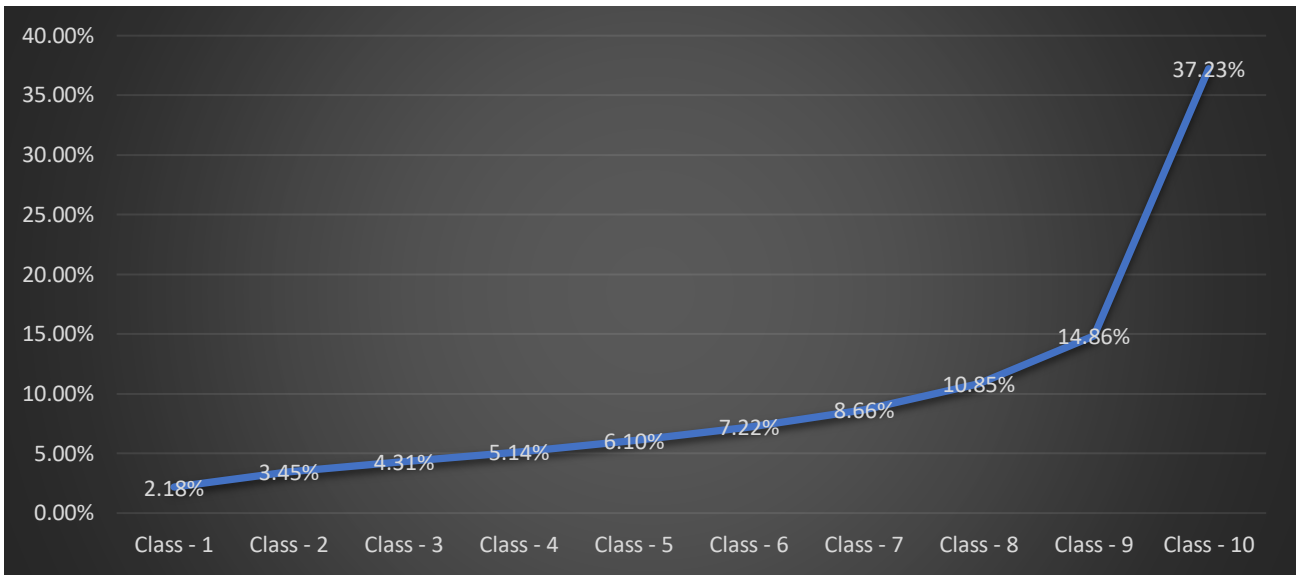

Figure 1. The Estimation of Income Class's Contribution (in percentage) to Income Decline Due to Carbon Tax Implementation in Indonesia

Overall, this study emphasizes that the impact of the carbon tax is felt by all income groups, but with a proportionally heavier burden on low-income households. In addition, urban households tend to be more affected nominally compared to rural households. The graph showing the difference in income before and after the implementation of the carbon tax clearly illustrates this nominal and proportional impact.

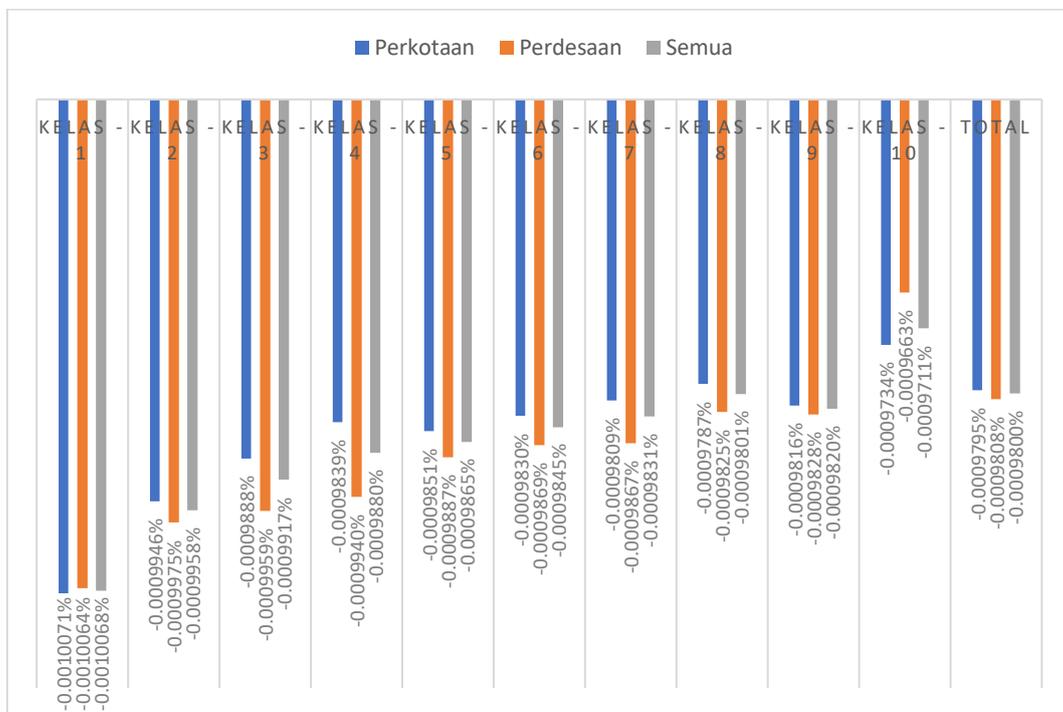

Figure 2. Estimation Percentage of Household Income Decrease Due to Carbon Tax Per Region and Income Class Based on Indonesia IO Table 2016

These results underscore the importance of designing more effective mitigation policies, especially to protect low-income households from the regressive impacts of carbon tax policies. This research suggests the need for additional policies such as energy subsidies or direct assistance to low-income households to reduce the burden imposed by these policies.

*The Impact of Carbon Tax on Income Distribution and Social Inequality in Indonesia*

The implementation of carbon tax also affects the pattern of income distribution. Estimation of the impact of carbon tax implementation on income distribution in Indonesia, both in urban areas (Urban), rural areas (Rural), and overall (Urban and Rural), is measured using the Gini Ratio and Lorenz Curve.

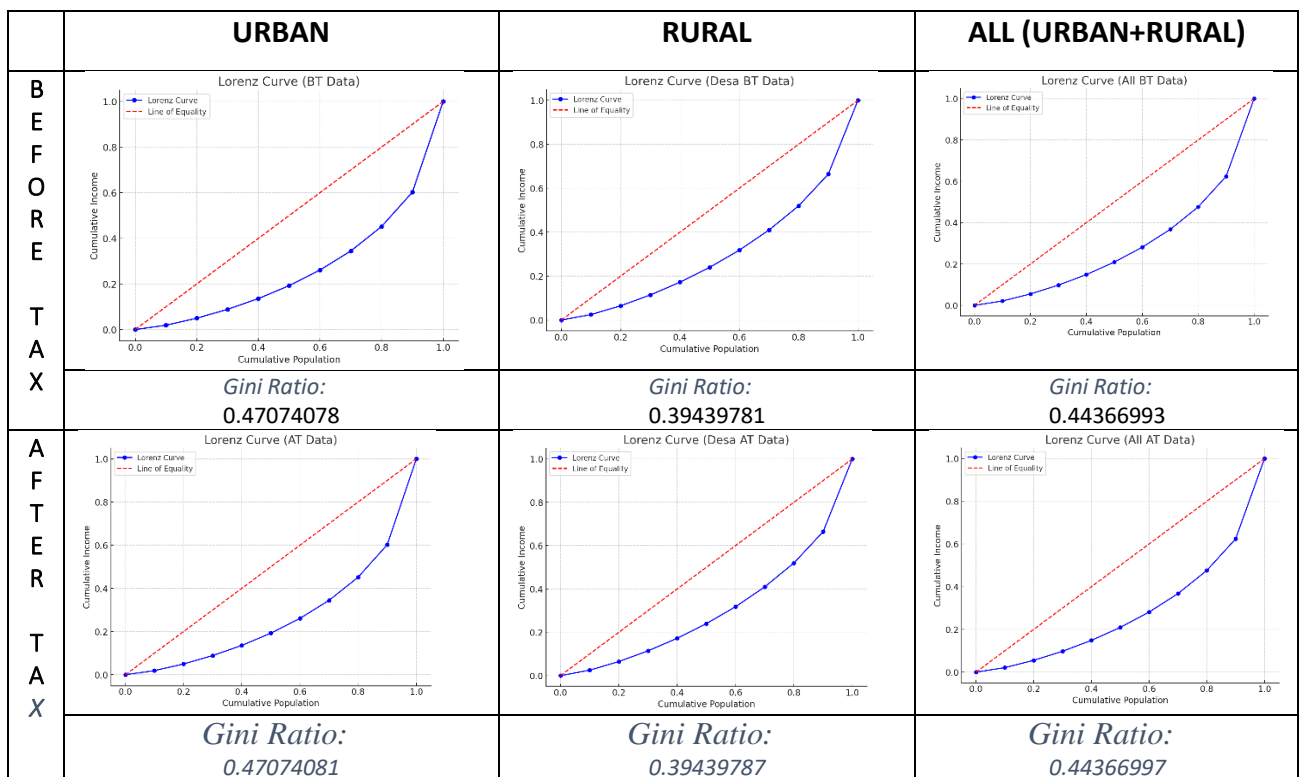

Figure 3. The Estimation Carbon Tax Implementation Impact on Income Distribution Both in Urban and Rural Areas in Indonesia Through Lorenz Curve and Gini Ratio Analysis

The Gini Ratio shows very little change in both urban and rural areas, as well as overall, after the implementation of the carbon tax. For urban areas, the Gini Ratio changed from 0.47074078 before the tax implementation to 0.47074081 after the tax implementation. In rural areas, the Gini Ratio changed from 0.39439781 to 0.39439787. Meanwhile, for the entire region, the Gini Ratio changed from 0.44366993 to 0.44366997.

Furthermore, the Lorenz Curve also shows that the pattern of income distribution changes very little after the implementation of the carbon tax. The line of perfect equality is shown by the red line on each graph, while the blue curve shows the actual income distribution. The changes indicated by the Gini Ratio and the Lorenz Curve are relatively small, but they can still be an indication of a tendency towards increasing inequality due to the implementation of the carbon tax.

The intensity of income changes, both absolutely and relatively, also needs to be considered. Although the overall impact may seem minimal, small changes in the Gini Ratio and income distribution patterns can indicate a greater impact on certain groups within society.

Thus, these results highlight the importance of formulating fairer and more effective policies to mitigate the negative impacts of carbon taxes. Mitigation policies such as subsidies or rebates for low-income households should be considered to prevent greater inequality. Further research is also needed to identify the groups most vulnerable to the impact of this policy and to formulate appropriate mitigation strategies.

**Discussion**

This research makes an important contribution to understanding the regressive impact of carbon tax, particularly in the context of Indonesia. Many countries have implemented carbon taxes, but these policies are often not strong enough to significantly reduce emissions because dependence on fossil fuels continues in various sectors (The World Bank, 2024a). The results of this study indicate that the implementation of carbon taxes in Indonesia affects all income classes, with low-income groups experiencing a proportionally greater impact. This reinforces the argument that without effective mitigation policies, carbon taxes cannot fully suppress dependence on fossil fuels, especially in developing countries. In line with these findings, Stern argues that carbon tax policies need to be accompanied by massive investments in renewable energy to reduce dependence on fossil-based energy (Stern, 2008).

Although carbon taxes aim to reduce carbon emissions, this policy often faces political and economic resistance, which hinders its implementation. This study shows that the nominal impact of carbon taxes is greater on high-income households, but the impact is felt more heavily by low-income households proportionally. This is in line with the findings of (Dallas Burtraw, 2009) which state that resistance to carbon taxes often stems from concerns that low-income groups will be disproportionately affected, leading to frequent rejection of such policies if adequate compensation measures are not in place. Therefore, it is important for policymakers in Indonesia to consider more effective mitigation strategies to address this resistance, such as direct cash assistance programs or renewable energy subsidies.

In developing countries like Indonesia, an economy structure that still heavily relies on the fossil-based energy sector slows down the transition to renewable energy. The results of this study confirm that urban households tend to be more nominally affected compared to rural households, indicating that the urban sector has higher energy consumption and, therefore, is more vulnerable to changes in energy prices due to carbon taxes. According to Borenstein and Bushnell (Borenstein & Bushnell, 2015), the transition to renewable energy requires strong policy support, including economic incentives for sectors that want to shift from fossil-based energy to green energy. These findings suggest that without adequate incentives, the energy transition in Indonesia could be hampered, given the importance of fossil-based energy in the country's economy.

One of the main challenges of carbon tax policy is its failure to address social injustice. The findings of this study indicate that low-income households are disproportionately affected, supporting the view that carbon taxes tend to be regressive if not accompanied by mitigation policies (Schmalensee & Stavins, 2017). Although the percentage decrease in income for high-income and low-income groups appears similar, the impact is much greater on the low-income group due to their limited purchasing power. Therefore, it is important for policymakers in Indonesia to design adequate compensation policies, such as energy subsidies or cash assistance to low-income households, to reduce the social inequality caused by the carbon tax policy.

Until now, the lack of in-depth analysis on how the carbon tax in Indonesia will impact household income distribution has hindered the formulation of appropriate mitigation policies. This research provides empirical data showing that the carbon tax indeed has a regressive impact, but it also highlights the need for additional policies to address this impact. As stated by Pizer and Sexton, in-depth analyses like this are crucial for designing fair and effective policies (Pizer & Sexton, 2019). The results of this research are expected to assist policymakers in Indonesia in formulating better mitigation strategies, to ensure that carbon tax policies are not only effective in reducing emissions but also fair for all layers of society.

To avoid all the gaps or issues that have been identified, the initial step that must be taken is to design a carbon tax policy considering the impact across all sectors of the economy, particularly on low-income households. The government must ensure that the transition to renewable energy is not only focused on reducing emissions but also on improving infrastructure and economic incentives that support the shift from fossil fuels to clean energy. Thus, dependence on fossil fuels can be gradually reduced without causing excessive resistance in the economic sectors that still rely on these resources.

Furthermore, it is important for policymakers to note that political and economic resistance can be overcome by providing adequate compensation to the most affected vulnerable groups. Subsidies for renewable energy, cash assistance, or income tax reductions can be effective solutions in mitigating resistance to carbon tax policies. With a more inclusive approach, the government can ensure that this policy runs effectively without triggering widespread resistance from the public or industry.

The solution has been identified through the results of this research, which highlights the importance of effective mitigation policies. The results of this study emphasize the importance of compensation for low-income households and the formulation of more inclusive policies to avoid regressive impacts. By implementing the solutions found in the Results section of this research, the government can ensure that carbon tax policies are not only effective in reducing carbon emissions but also fair and do not exacerbate social inequality.

This study concludes that the implementation of carbon tax in Indonesia is regressive, where low-income households bear a greater tax burden proportionally compared to high-income households. Although the percentage decrease in income appears similar across income groups, the nominal impact is felt more by the high-income group. However, the regressive impact is proportionally more significant on low-income groups due to their limited purchasing power. Therefore, mitigation policies such as energy subsidies or direct assistance are necessary to reduce the negative impact of this policy on vulnerable groups. Overall, this research emphasizes the importance of designing fair and inclusive policies in the implementation of carbon tax in Indonesia.

Theoretically, the findings of this research enrich the understanding of the regressive impact of carbon tax policies, particularly in developing countries like Indonesia. This research reinforces the literature stating that carbon tax policies can exacerbate income inequality if not accompanied by adequate compensation policies. From a practical standpoint, these findings provide clear recommendations to policymakers to consider the social impacts of carbon tax policies and develop more effective mitigation strategies. In the context of Macroeconomics and Sustainable Development, this research shows that fiscal policies designed to reduce emissions must consider their impact on low-income households to ensure social justice. These results can also encourage the formulation of more comprehensive and effective policies in supporting the transition towards a low-carbon economy without sacrificing social welfare.

Although this research provides valuable insights, there are several limitations that need to be acknowledged. One of them is that this study uses a quantitative approach with data limited to household income aspects. Other factors such as energy consumption behavior and the long-term impact of carbon tax policies have not been fully explored. Future research is recommended to delve deeper into the dynamic impacts of carbon taxes on specific sectors, as well as how these policies can be integrated with renewable energy policies. Further research also needs to involve more comprehensive distributional analysis to ensure that the implemented mitigation policies are truly effective in reducing regressive impacts across income groups.

**CONCLUSION**

This research provides a clear picture of the impact of carbon tax on household income distribution in Indonesia. The findings show that although this policy aims to reduce carbon emissions, its impact tends to be regressive, with low-income households bearing a disproportionately larger burden. Therefore, it is important for policymakers to consider mitigation mechanisms such as energy subsidies or direct assistance to alleviate the burden on vulnerable groups. With the implementation of more equitable policies, it is hoped that carbon taxes can serve as an effective tool in reducing emissions without exacerbating social inequality in Indonesia. This research emphasizes that the success of fiscal policies in supporting sustainable development must always align with the principles of social justice.

As a follow-up, further research is needed to explore other aspects that have not been thoroughly discussed in this study, such as the long-term impact of carbon tax policies on energy consumption behavior and their effects on certain sectors. Additionally, a more comprehensive study on the implementation of renewable energy policies needs to be conducted to support a faster and fairer transition towards a low-carbon economy. Thus, this research is expected to serve as a foundation for better discussions and policy formulation in the future, particularly in the context of inclusive and equitable sustainable development.

.

**Declaration of generative AI and AI-assisted technologies in the writing process**

During the preparation of this work the author(s) used [Quillbot] in order to [write in English correctly]. After using this tool/service, the author(s) reviewed and edited the content as needed and take(s) full responsibility for the content of the publication.